\documentclass[leqno]{article} 

\usepackage{latexsym}
\usepackage{amssymb}
\usepackage{amsthm}
\usepackage{amsmath}

\usepackage{graphics}

\theoremstyle{definition}

\renewcommand{\L}{\ensuremath{\mathrm{L}}}

\newcommand{\Cinf}{\ensuremath{\mathrm{C}^\infty}}
\newcommand{\D}{\ensuremath{{\cal D}}}
\renewcommand{\S}{\ensuremath{{\cal S}}}

\newcommand{\mb}[1]{\ensuremath{\mathbb{#1}}}
\newcommand{\N}{\mb{N}}

\newcommand{\R}{\mb{R}}

\newcommand{\WF}{\mathrm{WF}}
\newcommand{\singsupp}{\mathrm{sing supp}}

\renewcommand{\d}{\ensuremath{\partial}}
\newcommand{\diff}[1]{\frac{d}{d#1}}

\newfont{\bl}{msbm10 scaled \magstep2}

\newtheorem{thm}{Theorem}
\newtheorem{lemma}[thm]{Lemma}
\newtheorem{prop}[thm]{Proposition}

\newtheorem{rem}[thm]{Remark}

\newcommand{\emb}{\hookrightarrow}

\newcommand{\FT}[1]{\widehat{#1}}

\newcommand{\F}{\ensuremath{{\cal F}}}

\newcommand{\dis}[2]{\langle #1 , #2 \rangle}  
\newcommand{\inp}[2]{\langle #1 | #2 \rangle}  
\newcommand{\notmid}{\mid\kern-0.5em\not\kern0.5em}

\newcommand{\norm}[2]{{\| #1 \|}_{#2}}

\newcommand{\ltw}[1]{\norm{#1}{\L^2}}

\newcommand{\ga}{\gamma}
\newcommand{\de}{\delta}
\newcommand{\eps}{\varepsilon}

\newcommand{\vphi}{\varphi}

\newcommand{\supp}{\mathop{\mathrm{supp}}}

\newcommand{\ovl}[1]{\overline{#1}}

\newcommand{\mi}{\mbox{i}}
\newcommand{\msci}{\mbox{\scriptsize i}}

\newcommand{\hf}{{\textstyle{1\over2}}}

\newcommand{\fr}[2]{{ \displaystyle \frac{#1}{#2} }}

\newcommand{\beq}{\begin{eqnarray}}
\newcommand{\eeq}{\end{eqnarray}}
\newcommand{\beqa}{\begin{eqnarray}}
\newcommand{\eeqa}{\end{eqnarray}}

\newcommand{\ba}{\begin{array}}
\newcommand{\ea}{\end{array}}

\begin{document}
 
\title{Detection of wave front set perturbations via correlation:\\
       Foundation for wave-equation tomography} 
\author{G\"{u}nther H\"{o}rmann and Maarten V. de Hoop \\
        \textit{
        Department of Mathematical and Computer Sciences}, \\
        \textit{
        Colorado School of Mines, Golden CO 80401}}
\date{\today} 
\maketitle 
 
\begin{abstract} 
We discuss the mathematical aspects of wave field measurements used
in traveltime inversion from seismograms. The primary information about the 
medium is assumed to be carried by the wave front set and its perturbation 
with repsect to a hypothetical background medium is to be estimated. By
a convincing heuristics a detection procedure for this perturbation was
proposed based on optimization of wave field correlations. We investigate its
theoretical foundation in simple mathematical case studies using the 
distribution theoretic definition of oscillatory integrals.
\end{abstract} 

\section{Introduction}

In this paper, we investigate how to carry out tomography directly in
terms of wavefield measurements. Tomography, in its original form,
uses a `measured' wavefront set as input in an inversion procedure
which is solely (symplectic) geometric in nature, viz. based upon
finding bicharacteristics that result through a canonical relation in
matching the measurement. In `wave-equation' tomography, one aims at
replacing the geometric procedure by a wave-solution procedure, but
keeping the wavefront set of the measurements as the primary source of
information about the medium.

Following an embedding procedure to formulate the inverse problem,
i.e. introducing a background medium and incident field and a medium
contrast and scattered field, we then face the problem of detecting
perturbations in the wavefront set associated with the scattered
(perturbed $-$ incident) field. An intuitive choice is based upon
correlating the perturbed field with the incident field. We will show,
by example, that such procedure should be carried out delicately. In
fact, we conclude that the perturbation of the wavefront set can be
derived from the singular support (of the derivative) of the proposed
time correlation.

The outline of the paper is as follows. We briefly review the
microlocal representation of solutions to the scalar wave equation
(Section~2). In Section~3 we introduce the measuring process and its
mathematical implementation; we describe how the wavefront set of the
wavefield propagates through this measuring process. When we perturb
the coefficient function in the wave equation (the wave speed) the
solution representation will be perturbed. In particular, its
wavefront set will shift in the measurement-variables cotangent
bundle. We formulate the process of correlating, within the measuring
process, the perturbed representation with the original
representation, and identify how such shift appears in the result. It
is conjectured that the derivative of the (time) correlation at any
given measurement position has its singular support precisely at the
time shift associated with the perturbation of the wavefront set. In
Section~4 we give examples to illustrate the conjecture. Special
attention is paid how to define the product of distribution solutions
within the correlation process. Finally, in Section~5, we discuss a
method of detecting the singular support of the correlation in time at
any measuring position by means of `localized' Fourier transforms. The
procedure defines a criterion to develop wave-equation tomography.

\section{Fourier integral representation of wave \\
         solutions}


The scalar wave equation for acoustic waves in a constant density
medium is given by
\beq
   P u = f ,
\label{mod.1}
\eeq
with
\beq
   P = \partial_t^2 + \underbrace{D \ c(x)^2 D}_{A(x,D)} ,
\label{mod.2}
\eeq
where $D = -\mi \partial_x$. The equation is considered on an open
domain $\Omega \subset \R^n$ and in a time interval $]0,T[$.

We decouple the wave equation into its forward and backward
components. To this end, we introduce the elliptic operator $A(x,D)$
and its square root $B(x,D) = \sqrt{A(x,D)}$. Decomposing the field
according to
\beq
   u_{\pm} = \hf u \pm \hf \mi B(x,D)^{-1} \partial_t u ,
\label{mod.3}
\eeq
in combination with the source decomposition
\beq
   f_{\pm} = \pm \hf \mi B(x,D)^{-1} f ,
\label{mod.4}
\eeq
then results in the equivalent system of equations
\beq
   \left[ \partial_t \pm \mi B(x,D) \right] u_{\pm} = f_{\pm} .
\label{mod.5}
\eeq
Throughout, we assume that $c \in C^{\infty}(\Omega)$.  We will
construct operators $G_{\pm}$ with distribution kernels
$\mathcal{G}_{\pm}(x,x_0,t,t_0)$ that solve the initial value problem
equivalent to (\ref{mod.5}) with $f_{\pm} = \pm \delta$.

Let $H = H(x,\xi,\tau) = \tau \pm B^{\mathrm{prin}}(x,\xi)$ denote the
Hamiltonian either for the forward or backward wave propagation. The
Hamilton system of equations that generates the Hamiltonian flow or
bicharacteristics is given by
\beq
\ba{rclcrcl}
   \fr{\partial x}{\partial \lambda} &=&
    \pm \fr{\partial}{\partial \xi} B^{\mathrm{prin}} & \ ,\ &
   \fr{\partial t}{\partial \lambda} &=& 1 \ ,
\\ \\
   \fr{\partial \xi}{\partial \lambda}  &=&
    \mp \fr{\partial}{\partial x} B^{\mathrm{prin}} & \ ,\ &
   \fr{\partial \tau}{\partial \lambda} &=& 0 \ .
\ea
\label{mod.6}
\eeq
Observe that $H(x,\xi,\tau) = 0$ implies
 $\tau = \mp B^{\mathrm{prin}}(x,\xi)$.

Equation (\ref{mod.5}) can be solved, microlocally, in the form of a
Fourier integral representation. The phase of the associated Fourier
integral operator follows from the canonical relations
\[
   C_{\pm} = \{ (x(x_0,\xi_0,\pm t),t,\xi(x_0,\xi_0,\pm t),
                \underbrace{\mp B^{\mathrm{prin}}(x_0,\xi_0)}_{\tau}
             \ ;\ x_0,-\xi_0 \} .
\]
Let
\[
   (x_I,x_0,\underbrace{\xi_J,\tau}_{\theta})
\hspace*{0.5cm}\mbox{with}\hspace*{0.5cm}
   I \cup J = \{ 1,\ldots,n \}
\]
denote coordinates on $C_{\pm}$. A function $S$ will locally describe
$C_+$ according to
\beq
\ba{rclcrcl}
   x_J &=& -\fr{\partial}{\partial \xi_J} S & \ ,\ &
     t &=& -\fr{\partial}{\partial \tau} S \; ,
\\ \\
   \xi_I &=& \fr{\partial}{\partial x_I} S & \ ,\ &
   \xi_0 &=& -\fr{\partial}{\partial x_0} S \; ,
\ea
\label{mod.7}
\eeq      
and generates the non-degenerate phase function
\beq
   \phi_+(x,x_0,t,\xi_J,\tau) = S(x_I,x_0,\xi_J,\tau)
                          + \inp{\xi_J}{x_J} + \tau t .
\label{mod.8}
\eeq
In our notation, we will suppress the dependence on $x_0$ and collect
$\xi_J,\tau$ in the phase variables $\theta$. The canonical relation
can then be written as
\[
   C_+ = \{ ((x,t,\partial_x \phi_+,\partial_t \phi_+);
             (x_0,-\partial_{x_0} \phi_+)) \ |\
             \partial_{\theta} \phi_+ = 0 \} .
\]
We synthesize the canonical relation $C_{\phi} = C_+ \cup C_-$ with
associated (non-degenerate) phase function $\phi = \phi_-$ if $\tau >
0$, $\phi = \phi_+$ if $\tau < 0$. In accordance with (\ref{mod.3}) we
obtain
\beq
   G(x,x_0,t) = \hf \mi [ G_+(x,x_0,t) - G_-(x,x_0,t) ] \,
                B(x_0,D_{x_0})^{-1} .
\label{mod.9}
\eeq
With this fundamental solution, the solution of (\ref{mod.5}) and its
dependence on the initial conditions can then be written in the form
of a Fourier integral operator (FIO) with amplitude $a =
a(x_I,x_0,\xi_J,\tau)$. In fact, $a$ is a section of the tensor product
$M_{C_{\phi}} \otimes \Omega^{1/2}(C_{\phi})$ of the Keller-Maslov
line bundle and the half-densities on $C_M$.

The kernel of the FIO admits an oscillatory integral (OI)
representation. In the remainder of this paper we consider such OIs to
represent `the wavefield'. Perturbation of this wavefield are induced
by perturbation of the coefficient function $c(x)$.

\section{Detection of singularities of the wave field}

As described above, each component of the wave field as well as the
perturbed wave field can be represented by an OI,
\begin{equation}\label{osc_int}
   u(x,t) =  \int e^{\msci \phi(x,t,\theta)} a(x,t,\theta)  \, d\theta	,
\end{equation}
where $\phi$ is a non-degenerate phase function and $a$ a symbol
(\cite{Hoermander:V1}, Sect.~7.8); note that the wave front set
satisfies the inclusion (\cite{Hoermander:V1}, Thm.~8.1.9)
\begin{equation}\label{wf_osc}
   \WF(u) \subseteq \{(x,t;\d_x\phi(x,t,\theta),\d_t\phi(x,t,\theta))
   \mid \d_\theta\phi(x,t,\theta) = 0 \} .
\end{equation}

\subsection{Measurements as restrictions to submanifolds}

Measurements are recordings of the wave field $u$ in stations at
certain points $x$ in the acquisition manifold over some time interval
$(t_0,t_1)$; mathematically, this corresponds to the restriction of
the distribution $u$ to the one-dimensional submanifolds $S_x = \{ x\}
\times \R$ followed by further restriction of the resulting
one-dimensional distribution $u_x$ of time to the open interval
$(t_0,t_1)$.

While the second of those restrictions is always possible and
straightforward, the first can be carried out as continuous map only
on distributions satisfying the following condition
(\cite{Hoermander:V1}, Thm.~8.2.4 and Cor.~8.2.7)
\begin{equation}\label{res_cond}
   \WF(u) \cap \{(x,t;\xi,0) \mid t \in \R,\ \xi \in \R^n \}
   = \emptyset .
\end{equation}
Note that by (\ref{wf_osc}) this condition is satisfied if and only if
$\d_t \phi(x,t,\theta) \not = 0$ whenever $\d_\theta\phi(x,t,\theta) =
0$.  If it holds, the restriction $u_x$ can be defined as the pullback
$\iota_x^* u$ of $u$ under the embedding map $\iota_x : S_x \emb
\R^{n+1}$ and by (\ref{wf_osc}) we have the wave front set relation
\begin{multline}\label{wf_res}
   \WF(u_x) \! \subseteq \{ (t;\tau) \mid \exists\xi: 
                            (x,t;\xi,\tau) \in \WF(u) \} \\
    \subseteq \{ (t;\d_t \phi(x,t,\theta)) \mid
                   \d_\theta \phi(x,t,\theta) = 0 \} .
\end{multline}

Let $\psi$ be a phase function, $b$ a symbol, both with the same
domains and supports as $\phi$, $a$, and $v$ be the oscillatory
integral defined by them; assume that $v$ also satisfies
(\ref{res_cond}) and set $v_x = \iota_x^* v$. In case we are
interested only in a certain time window of measurement we may use
further cut-offs and achieve that $u_x$ and $v_x$ are compactly
supported.

\subsection{The correlation function}

For $a\in\R$ denote by $T_a$ the translation by $a$ on $\R$. If the
distributional product $w_{x,t} = u_x \cdot \ovl{T_t^* v_x}$ can be
defined and yields an integrable distribution (\cite{Horvath:66},
Sect.~4.5) we define the value of the \emph{correlation function} at
$t$ by
\begin{equation}\label{corr}
   c[u_x,v_x](t)
   = \dis{u_x \cdot \ovl{T_t^* v_x}}{1} = \dis{w_{x,t}}{1} .
\end{equation}
The correlation is bilinear in $[.,.]$. Whenever there is no
ambiguity about the distributions $u$ and $v$ and the point $x$ under
consideration we will denote the correlation briefly by $c(t)$.

Whenever $u$ and $v$ represent the unperturbed and perturbed solution,
then typically $\d_t \phi = \d_t \psi$ (the frequencies coincide) and
therefore for certain values of $t$ we expect the cotangent components
of the wave front sets of $u_x$ and $T_t v_x$ to be identical on the
overlap of singular supports. That means that, unless both cotangent
parts are only half rays on the same side of $0$, H\"{o}rmander's
condition (\cite{Hoermander:V1}, Thm.~8.2.1) for defining the product
does not apply. But within the hierarchy of distributional products
described by Oberguggenberger (\cite{O:92}, Ch.~II) this condition,
`WF favorable', appears only as one out of a variety of consistent
possibilities to give a distributional meaning to the product under
consideration. We apply some of these to the analysis of the
correlation function in some examples below to explore and illustrate
whether and how the correlation, after restriction, can provide
information about shifts in wave front set from $u$ to $v$. It will
become clear that the customary criterion of searching for the
`stationary point' of the correlation (Dahlen, Hung and
Nolet~\cite{Da:00}, Zhao, Jordan and Chapman~\cite{Zh:00} and Luo and
Schuster~\cite{LS:91}) for detecting the shift in wave front sets is
generally incorrect.

Here, we would like to point out that the appropriate mathematical
framework to deal with the multiplication (and also the
restrictability) in a uniform and systematic manner is Colombeau's
theory of generalized functions (cf.~\cite{Colombeau:85,O:92}). Such
framework will enable us to cope with the integrability question
(forming the correlation) at the same time (\cite{Hoermann:99}).

%
%

Practically, we will have to consider regularizations or
approximations to the formal expression $c(t) = \dis{w_{x,t}}{1}$ of
the correlation either to give a meaning to the product or to make the
integration (i.e., distributional action on $1$) well-defined. This
amounts to the attempt of defining $c(t)$ as the pointwise (in $t$)
limit of sequences
\[
        c_n(t) = \dis{w^n_{x,t}}{1}
\]
as $n\to\infty$ where $w^n_{x,t}$ is a suitable regularization or
approximation of $w_{x,t}$.

\subsection{The shift of singular supports}

We compare the singular supports, or rather the wave front sets, of
$u_x$ and $v_x$. Their offset expresses the amount of time shift of the wave
fronts (or rather singularities) at location $x$ by the perturbation.

First we observe that under a natural time evolution condition on the
phase function a restrictable OI is representable as an OI in one
dimension.

\begin{lemma} If
$u = \int a(.,\theta) e^{\msci \phi(.,\theta)} \, d\theta
\in \D'(\R^{n+1})$ satisfies condition (\ref{res_cond}) at $x$ and
$\d_t\phi(x,t,\theta) \not= 0$ for all $t$ and $\theta \not=0$ such
that $(x,t,\theta)\in \supp(a)$ then the restriction $u_x$ to $S_x$ is
the OI on $\R$ (i.e., in the time variable) where $x$ is considered as
a parameter in the phase and amplitude. Therefore
\begin{equation}\label{ux_oi}
   u_x = \int a(x,.,\theta) e^{\msci \phi(x,.,\theta)} \, d\theta .
\end{equation}
\end{lemma}
\begin{proof}
By assumption $\phi_x(t,\theta)=\phi(x,t,\theta)$ defines a phase
function on $\R\times\R^N$. We have $u_x = \iota_x^*(u)$ and
$\iota_x^*$ is continuous on the subspace of restrictable
distributions. Therefore we may use any standard OI regularization $u
= \lim_{\eps\to 0} u_\eps$ and obtain $u_x = \lim_{\eps\to 0}
\iota_x^*(u_\eps)$. Since the latter is an OI regularization in one
dimension with phase function $\phi_x$ and symbol $a(x,.,.)$ the
assertion is proved.
\end{proof}

Note that the usual stationary phase argument applied to this
one-dimensional OI gives the same upper bound for the wave front set
as established above in (\ref{wf_res}). Assuming that the perturbed
solution $v$ is given as an OI with phase function $\psi$ and
amplitude $b$ we can compare the wave front sets of their measurements
at $x$ (restrictions to $S_x$).

As pointed out above, the perturbation will affect the phase function
only in its $x$- and $\theta$-gradient, i.e., we may assume that $\d_t
\phi = \d_t \psi$. If $(t_0,\tau_0) \in \WF(u_x)$ then $\tau_0 = \d_t
\phi(x,t_0,\theta_0)$ for some $\theta_0$ with $\d_\theta
\phi(x,t_0,\theta_0) = 0$; similarly if $(t_1,\tau_1)\in\WF(v_x)$ then
$\tau_1 = \d_t \phi(x,t_1,\theta_1)$ for some $\theta_1$ with
$\d_\theta \phi(x,t_1,\theta_1) = 0$. In any microlocal representation
of the solution to the wave equation, in the absence of attenuation,
the phase contains $t$ only linearly, say, in the form $t \rho$ for some
conjugate (frequency) variable $\rho$.

As was shown in Section~1, typical phase functions are of the special
form $\phi(x,t,\eta,\rho) = \phi_0(x,\eta,\rho) - t \rho$ and
$\psi(x,t,\eta,\rho) = \psi_0(x,\eta,\rho) - t \rho$. In this case the
stationary phase conditions (see eq.~(\ref{wf_osc})) in the wave front sets 
read
\begin{eqnarray}
   t_0 = \d_\rho \phi_0(x,\eta_0,\rho_0), \qquad 
         \d_\eta \phi_0(x,\eta_0,\rho_0) = 0  \\
   t_1 = \d_\rho \psi_0(x,\eta_1,\rho_1), \qquad 
         \d_\eta \psi_0(x,\eta_1,\rho_1) = 0     
\end{eqnarray}
and the respective $t$-derivatives of the phases yield cotangent
components $\tau_0 = - \rho_0$ and $\tau_1 = - \rho_1$. By the
(positive) homogeneity of $\phi$ and $\psi$ w.r.t.\ $(\eta,\rho)$
their first order derivatives w.r.t.\ those variables are (positively)
homogeneous of degree $0$.

Hence, if we are detecting time-like forward (resp.\ backward)
cotangent directions, i.e., $\rho > 0$ (resp.\ $\rho < 0$), we may
rescale the arguments in the phase and obtain the time shifts
\begin{equation}
   t_1 - t_0 = \d_\rho \psi_0(x,\eta'_1,\pm 1) - 
               \d_\rho \phi_0(x,\eta'_0,\pm 1) 
\end{equation}
for the corresponding (slowness co-vector) projections $\eta'_0$,
$\eta'_1$ satisfying the conditions
\begin{equation}
   \d_\eta \phi_0(x,\eta'_0,\pm 1) = 0, \qquad 
   \d_\eta \psi_0(x,\eta'_1,\pm 1) = 0 .
\end{equation}

\subsection{Correlation optimization}

In \cite{LS:91} a traveltime inversion method is described that uses
optimal fitting of traveltimes from synthetic seismograms according to
wave equation solutions of velocity model perturbations. The fitting
criterion is based upon a crosscorrelation function of the observed
($v$) and the synthetic ($u$) seismic data. This crosscorrelation of
\cite{LS:91} corresponds to the correlation function defined in
(\ref{corr}) above.

We give a brief schematic description of this interesting fitting
strategy and test its theoretical validity in three simple examples
below. Assume that $v$ represents the observed (or perturbed) wave
field and $u = u[\ga]$ is the solution of a velocity model which is
parametrized by the variable velocity $\ga(x)$. We assume that $\ga$
is a real-valued smooth function. Therefore the correlation function
is actually dependent on time $t$ and the velocity $\ga$ which we
indicate in the notation
\[
   c(t)[\ga] = \dis{u_x[\ga] \cdot \ovl{T_t^* v_x}}{1} ,
\]
where $(.)$ denotes the scalar and $[.]$ the functional argument of
$c$. An intuitive expectation would then be that at the exact
traveltime shift induced by the perturbation, we find optimum match
(overlap) of the corresponding seismograms and therefore the
crosscorrelation should be maximal. Leaving possible maxima at time
interval boundaries aside, we search for a $(\ga,t)$ relation that
gives stationarity of the crosscorrelation, i.e.,
\begin{equation}\label{corr_stat}
   F(t)[\ga] = \d_t c(t)[\ga] \equiv 0 .
\end{equation}

Naively speaking we can consider this to be an implicit definition of
a functional relationship between $\ga$ and $t$.
(Observe that $\ga$ is an infinite-dimensional variable and therefore 
more attention is to be paid to the exact meaning of applying an `implicit
function theorem' below.) 
Under the condition that $\d_t F = \d_t^2 c \not=
0$ we would therefore try to solve equation (\ref{corr_stat}) locally
for $t$ as a function of $\ga$ and find a quasi-explicit
representation by
\[
   \d_\ga t = - \frac{\d_\ga F}{\d_t F} .
\]

\section{Case studies}

\subsection{Two propagating delta waves}

Consider $u = \de_0(x-s)$ and $v = \de_0(x - \ga s)$, two Dirac deltas
travelling along the lines $x = s$ and $x = \ga s$ respectively.
(These are distributional pullbacks of $\de_0 \in \D'(\R)$, the Dirac
measure located at $0$, via the maps $(x,s)\mapsto x - s$ and $(x,s)
\mapsto x - \ga s$.) Assume that $x > 0$; the opposite sign case is
completely symmetric. We clearly have $u_x = \de_x$ and $\ovl{T^*_t
v_x} = \frac{1}{\ga}\de_{\frac{x}{\ga} - t}$, and therefore
\[
   \singsupp(u_x) = \{ x \}, \qquad
   \singsupp(v_x) = \{ \frac{x}{\ga} \} 
\]
yielding a singularity shift of $t_1 - t_0 = -x(1-1/\ga)$.

Observe that $u_x$ and $\ovl{T^*_t v_x}$ have disjoint singular
supports unless $t = -x(1-1/\ga)$ in which case their product would
require to multiply $\de_x$ with itself. This cannot be done
consistently within the hierarchy of distributional products (cf.\
\cite{O:92}) and calls for a systematic treatment in the framework of
algebras of generalized functions. However, here we touch upon those
aspects only in terms of regularizations.

Choose a rapidly decaying smooth function $\rho$ on $\R$ such that
$\int \rho = 1$, in other words $\rho$ is a mollifier, and set
$\rho_\eps(s) = \rho(s/\eps)/\eps$. Denote by $u_x^\eps$ and
$v_x^\eps$ the convolutions of $u_x$ and $v_x$ with $\rho_\eps$. Then
we have
\[
   u_x^\eps(s) = \rho_\eps(s-x),\qquad \ovl{T_t^* v_x^\eps}(s) =
                 \frac{1}{\ga}\ovl{\rho_\eps(s + t - x/\ga)}
\]
and upon integration of $u_x^\eps(s)\ovl{T_t^* v_x^\eps}(s)$ w.r.t.\
$s$ with a change of the variable $y = (s-x)/\eps$ we obtain for the
regularized correlation function
\begin{equation}\label{reg_corr}
   c_\eps(t) = \frac{1}{\ga \eps} \int \rho(s)
    \ovl{\rho\big( s + \frac{t+x(1-1/\ga)}{\eps} \big)} \, ds . 
\end{equation}

If we let $\eps \to 0$ we observe that $c_\eps(t) \to 0$ pointwise for
$t \not= \bar{t} := -x(1-1/\ga)$ and $|c_\eps(\bar{t})| \to \infty$. 
Hence,
in an approximative sense, the singular support of the correlation
$c(t)$ contains the time shift information.  To be more precise, it is
not difficult to show that in the sense of distributions
\begin{equation}\label{corr1}
   c_\eps \to \frac{1}{\ga} \de_{-x(1-1/\ga)} . 
\end{equation}
For this, we just note that for arbitrary $\vphi\in\D(\R)$ one may
change the variable in $\int \vphi(t) c_\eps(t)\, dt$ to $r =
(t-\bar{t})/\eps$ and use the fact that $\int f*g = \int f \cdot
\int g$ for rapidly decreasing functions $f$ and $g$.

In particular, this shows that here the correlation is
stable under changes within the chosen class of regularizations since the
limit does not depend on $\rho$.

Curiously enough, the regularization approach also gives the correct
answer when using the procedure of \cite{LS:91}. Define the short-hand
notation $k_\eps(x,t,\ga) = (t-\bar{t})/\eps$ and consider
\[
   c_\eps'(t) = \frac{1}{\ga\eps^2} \int \rho(s) 
                \ovl{\rho'(s + k_\eps(x,t,\ga))} \, ds  
\]
and set $F_\eps(x,t,\ga) = \ga \eps^2 c_\eps'(t)$. We see that
\[
   \d_t \eps F_\eps(x,t,\ga)
   = \int \rho(s) \ovl{\rho''(s + k_\eps(x,t,\ga))} \, ds ,
\]
which is proportional to $\ltw{\rho'}^2$ at $t = \bar{t}$ and stays nonzero 
for all $t$ and $\gamma$ close enough. This in particular true at 
$\gamma = 1$ in which case $\bar{t} =0$. Therefore, in a
(possibly smaller) neighborhood of these values for $t$ and $\gamma$ we can 
solve the implicit equation $F_\eps(x,t,\ga) = 0$ for $t = t(x,\ga)$ and 
find locally
\[
   \d_\ga t(x,\ga)
   = - \frac{\d_\ga F_\eps(x,t,\ga)}{\d_t F_\eps(x,t,\ga)}
   = - \frac{\frac{x}{\ga^2\eps}
       \int \rho(s) \ovl{\rho''(s + k_\eps(x,t,\ga))}\,ds}
    {\frac{1}{\eps}\int \rho(s) \ovl{\rho''(s + k_\eps(x,t,\ga))}\,ds} 
   = - \frac{x}{\ga^2} . 
\]
We find from this by integration over $\ga$ (close to $1$) that
\[
        t(x,\ga) = \frac{x}{\ga} - x = -x (1-1/\ga)
\]
which is the correct shift of the singular support.

\subsection{A delta wave interacting with a shock}

We set $u = \de_0(x-t)$ and $v = H(x - \ga t)$ (where $H$ is the
Heaviside function) yielding exactly the same configuration of wave
front sets as in the previous case. In this case, restricting our
attention again to $x > 0$, $u_x = \de_x$ and $\ovl{T_t^* v_x}(s) =
H(x - \ga (s+t))$. The only critical product appears if $t = \bar{t} =
-x(1-1/\ga)$: At this point we have to deal with $\de_x(s) \cdot
H(x-s)$ which exists as a so-called `strict product (7.4)' in the
notion of \cite{O:92}, Ch.~II, assigning the value $\hf \de_x$ to
it. For $t < \bar{t}$ we obtain $u_x \cdot \ovl{T_t^* v_x} = \de_x$
and for $t > \bar{t}$ we have $u_x \cdot \ovl{T_t^* v_x} = 0$ because
the Heaviside contribution is constant $0$ or $1$ in those regions. In
summary
\[
   w_{x,t} = u_x \cdot \ovl{T_t^* v_x}
   = \begin{cases}
     \de_x & \text{if } t < \bar{t} \\
     \frac{1}{2} \de_x & \text{if } t = \bar{t} \\
     0 & \text{if } t > \bar{t}
     \end{cases} .
\]

If we interpret $\dis{w_{x,t}}{1}$ via the Fourier transform of
$w_{x,t}$ as $\FT{w_{x,t}}(0) = c(t)$ then we obtain $c(t)$ as the
measurable function
\[
   c(t) = \begin{cases}
          1 & \text{if } t < \bar{t} \\
          \frac{1}{2}  & \text{if } t = \bar{t} \\
          0 & \text{if } t > \bar{t}
          \end{cases} .
\]
As in the previous case, we observe that it is exactly the singular
support of $c$ -- here, the point $\bar{t} = -x(1-1/\ga)$ -- that
reveals the information of the correct shift.

Observe, however, that the travel time $\bar{t}$ is in fact the only
point where the (distributional) derivative $c'(t) = \de_{\bar{t}}$
does \textit{not} vanish. The previous evaluation based upon the
implicit function theorem hence does not apply.

\subsection{Wave equations with different medium constants}

We now return to the wave equation (Section~2), assume constant
coefficients, and invoke an exact solution representation rather than
an asymptotic one. We consider propagation in one spatial dimension.

Let $\chi \in \Cinf(\R)$ be real-valued, $\chi(-\xi) = \chi(\xi)$,
$\chi \equiv 0$ in a neighborhood of $0$ and $\chi \equiv 1$ for
$|\xi| \geq 1$; $\ga$ a constant $> 0$.

In the sense of OIs
\begin{eqnarray*}
   u(x,t) &=&
   \int e^{\msci ( t |\xi| - x \xi)}\,
        \frac{\chi(\xi)}{|\xi|} \, d\xi \\
   v(x,t) &=&
   \int e^{\msci( t |\xi| - x \xi/\ga)} \, 
        \frac{\chi(\xi/\ga)}{\ga|\xi|} \, d\xi
\end{eqnarray*}
($u$ respectively $v$ are the complex conjugates of $2\pi \mi$ times
the subtrahends in OI representation of the fundamental solutions for
the d'Alembert operator with wavespeed equal to $1$ and $\ga$,
respectively.)

The general $\WF$-bounds according to (\ref{wf_osc}) give
\begin{eqnarray*}
   \WF(u) &\subseteq&
   \{ (t,\pm t,-\xi,|\xi|) \mid t\in\R, \pm\xi > 0 \} \\
   \WF(v) &\subseteq&
   \{ (\ga t,\pm t,-\xi/\ga,|\xi|) \mid t\in\R, \pm\xi > 0 \} .
\end{eqnarray*}
Observe that half rays in cotangent components are minimal closed
cones in $\R^2\setminus 0$. We will show that, in fact, the inclusion
should be replaced by equality.

For symmetry reasons, we give detailed arguments in quadrant $x > 0$,
$t > 0$ only. Since $(\d^2_t - \d^2_x) u = 0$ and $(\d^2_t - \ga^2
\d^2_x) v = 0$, the theorem on the propagation of singularities
\cite{Hoermander:V1}, Thm.~8.3.3, applies; in particular, if $(t,t)\in
\singsupp(u)$ (resp.\ $(\ga t,t)\in\singsupp(v)$) then the whole line
through this point with directional vector $(1,1)$ (resp.\ $(\ga,1)$)
is in the singular support (with the same perpendicular cotangent
component in the wave front set attached to it). Therefore, to prove
equality in the above $\WF$ inclusion relations, it suffices to show
that $u$ (resp.\ $v$) is not smooth near $(0,0)$

Assuming the contrary, would imply that the function $x \mapsto \d_t
u(x,0)$ is smooth; but it is also equal to the Fourier transform of
$\mi \chi$, which cannot be smooth since $\F\chi = \F(\chi -1 + 1) =
\F(\chi - 1) + 2\pi \de$ where the first term is a smooth function (of
rapid decay) since $\chi -1$ is smooth and of compact support. The
argument for $\d_t v(x,0)$ is the same. We conclude that
\begin{eqnarray}
   \WF(u) &=& \{ (t,\pm t,-\xi,|\xi|) \mid t\in\R, \pm\xi > 0 \} \\
   \WF(v) &=&
   \{ (\ga t,\pm t,-\xi/\ga,|\xi|) \mid t\in\R, \pm\xi > 0 \} .
\end{eqnarray}

It follows immediately that both $u$ and $v$ are restrictable to $S_x$
and $u_x$ and $v_x$ are represented as the one-dimensional OIs where
$x$ appears as parameter in the phase only. Assuming $x > 0$, we
clearly have
\begin{eqnarray}
   \WF(u_x) &=& \{ (\pm x,\xi) \mid  \xi > 0 \} \\
   \WF(v_x) &=& \{ (\pm x/\ga,\xi) \mid \xi > 0 \} .
\end{eqnarray}
But then the time shift is given by
\begin{equation}
   t_1 - t_0 = \mp x(1 - 1/\ga) ,
\end{equation}
as expected from physical intuition.

In the remainder of this section we analyze the correlation in detail,
and investigate how the time shift appears. In the correlation we have
to multiply the distributions
\[
   u_x(s) = \int e^{\msci (s|\xi| - x\xi)}
            \frac{\chi(\xi)}{|\xi|} \, d\xi
\]
and 
\[
   \ovl{T_t v_x(s)}
          = \int e^{-\msci ((s+t)|\xi| - x\xi/\ga)}
            \frac{\chi(\xi/\ga)}{\ga|\xi|}\, d\xi ,
\]
which have wave front sets
\begin{eqnarray*}
   \WF(u_x) &=& \{-x,x\}\times \R_+ \\
   \WF(\ovl{T_t v_x})
            &=& \{-\frac{x}{\ga} - t, \frac{x}{\ga} - t\} 
                \times \R_+ .
\end{eqnarray*}
Hence, whenever $t \not= \mp x(1 \mp 1/\ga)$, the distributions have
disjoint singular supports and in case $t = \mp x(1 \mp 1/\ga)$ the
cotangent vectors in their wave front set cannot add up to $0$. We
conclude that for all $t$ the wave front sets are in favorable
position and the product $w_{x,t} = u_x \cdot \ovl{T_t v_x} \in
\D'(\R)$ can be defined in the sense of \cite{Hoermander:V1},
Thm.~8.2.10. The following lemma states that we are even allowed to
use the naive product of the OI expressions.

\begin{lemma} $w_{x,t}$ is (essentially) an OI given by
\begin{equation}\label{w_OI}
   w_{x,t} (s) = \int\!\!\!\int
      e^{\msci\big( s(|\xi| - |\eta|) - t|\eta| - x(\xi - \eta/\ga)
              \big)}
      \frac{\chi(\xi) \chi(\eta/\ga)}{\ga |\xi| |\eta|} \, d\xi d\eta
\end{equation}
and $t \mapsto w_{x,t}$ is weakly continuous $\R \to \D'(\R)$. We
introduce the following notation:
\begin{eqnarray*}
   \phi_t(s;\xi,\eta)
   &=& s(|\xi| - |\eta|) - t|\eta| -x(\xi - \eta/\ga) \\
   a(\xi,\eta)
   &=& \frac{\chi(\xi) \chi(\eta/\ga)}{\ga |\xi| |\eta|} 
\end{eqnarray*}
for the phase function and the amplitude.
\end{lemma}
\begin{proof}
For the justification of (\ref{w_OI}) we use the construction of the
distributional product in \cite{Hoermander:V1}, Thm.~8.2.10 via the
pullback of the tensor product on $\R^2$ under the map $\iota(s) =
(s,s)$ which embeds $\R$ as the diagonal into $\R^2$. In doing so the
original OIs may be approximated by smooth regularizations (e.g.,
amplitude cut-offs in the integrands) the tensor products thereof
being pulled back simply as smooth functions (meaning restriction to
$(s,s)$ in this case).

It is easily seen then that the smooth functions obtained thereby
converge weakly (as OI regularizations) to the OI given in
(\ref{w_OI}). By continuity of the pullback (under the given wave
front set conditions) this limit equals the pullback of the tensor
product of the corresponding limits and therefore, in turn, is the
distributional product $w_{x,t} = u_x \cdot \ovl{T_t v_x}$.

Note that $a$ is smooth in $(\xi,\eta)$ (due to the cut-off $\chi$)
and homogeneous of degree $-2$ outside the set $\{ |\xi| \geq 1,
|\eta| \geq \ga\}$ and is therefore a symbol of order $-2$. The
function $\phi_t$ is smooth on $\supp(a)$ and homogeneous of degree
$1$ in $(\xi,\eta)$. If $|t| \not= |x(1 \pm 1/\ga)|$ then the gradient
$\d_{(s,\xi,\eta)} \phi_t \not= (0,0,0)$ for all $(s,\xi,\eta)$ and
hence $\phi_t$ is a phase function.

In case $|t| = |x(1 \pm 1/\ga)|$ the gradient vanishes exactly along
one half-ray component of the set $\{ (\xi,\eta) \mid |\xi| = |\eta|
\}$ (e.g., along $\xi = \eta > 0$ if $t = -x(1-1/\ga)$). Although it
is no longer a phase function in the strict sense, the distribution
$w_{x,t}$ is then defined as the sum of a classical integral, an OI,
and a Fourier transform of an $\L^2$-function. We discuss this for the
case $t= -x(1-1/\ga)$ in detail, the other cases are completely
analogous.

Let $\mu(\xi,\eta)$ be a smooth function that is equal to $1$ near
$\xi = \eta > 1$, has support in $\{ \xi > 0, \eta > 0 \}$, and
satisfies $0 \leq \mu \leq 1$. Let $\nu(\xi,\eta)$ be smooth with
compact support and $\nu(\xi,\eta) = 1$ when $\xi^2 + \eta^2 \leq 1$.

\centerline{\includegraphics*{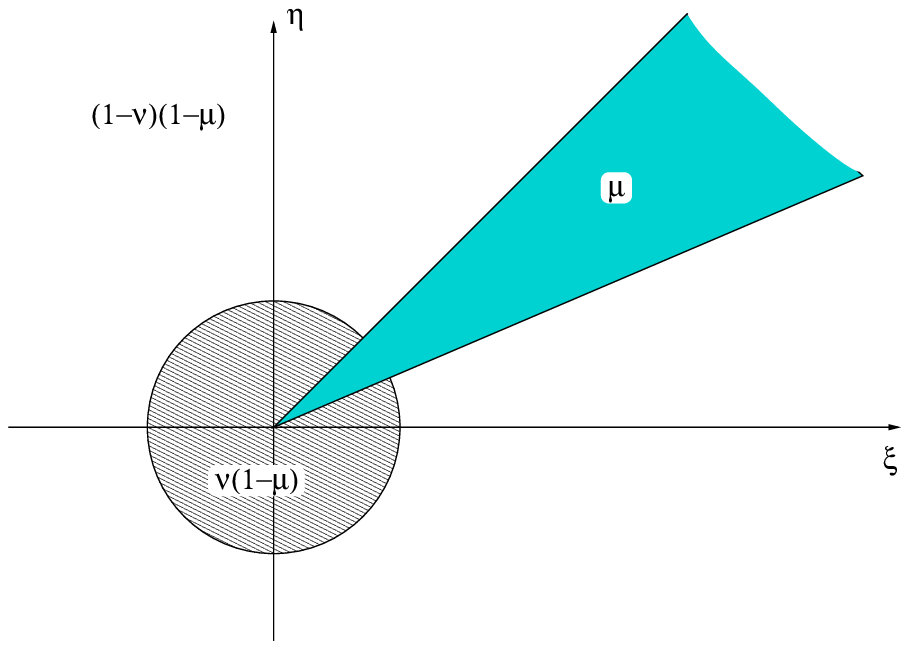}}

We can split the integral defining $w_{x,t}$ into three terms
according to $1 = \nu (1-\mu) + (1-\nu)(1-\mu) + \mu$. The first
integral, then, is a classical one defining a smooth function, the
second is an OI since the gradient of $\phi_t$ does not vanish on the
support of the integrand. In the third integral, we have
\[
   \phi_t(s;\xi,\eta)
   = (s-x)(\xi - \eta) = - \inp{(x-s,s-x)}{(\xi,\eta)}
\]
(insert $t = -x(1-1/\ga)$ and use the fact that $|\xi| = \xi > 0$ and
$|\eta| = \eta > 0$ on the support of the integrand) and hence the
last term is equal to
\[
   \int e^{-\msci \inp{(x-s,s-x)}{(\xi,\eta)}}
        \mu(\xi,\eta) a(\xi,\eta)\, d\xi d\eta ,
\]
which we interpret via the Fourier transform of the $\L^2$-function
$\mu a$ on $\R^2$ as $s \mapsto \F(\mu a)(x-s,s-x)$ in the sense of
locally integrable functions --- hence it is distribution on $\R$.

The weak continuity w.r.t.\ $t$ follows from the smooth dependence of
the phase function in the OI representation (cf.
\cite{Duistermaat:96}, before Thm.~2.2.2) and the continuity of the
Fourier transform on $\L^2$.
\end{proof}

\begin{rem}\label{weak_smooth_remark} From the last part of the proof
it follows that $t \mapsto w_{x,t}$ is weakly smooth on $\R \setminus
\{\pm x(1 \pm 1/\ga)\}$.
\end{rem}

In order to define the correlation function, we need to check whether
the action of $w_{x,t}$ on $1$ is well defined. We will do so by
showing that $w_{x,t}$ is tempered with Fourier transform
$\FT{w_{x,t}}$ being in fact a continuous function. This function can
be evaluated at $0$ yielding the interpretation $\dis{w_{x,t}}{1} =
\FT{w_{x,t}}(0)$.

We use an OI regularization of $w_{x,t}$ via the symmetric cut-off
function $\rho(\xi,\eta) = \rho_0(\xi) \rho_0(\eta)$ where
$\rho_0\in\D(\R)$ with $\rho_0(r) = 0$ when $|r| \geq 1$, $\rho_0(r) =
1$ when $|r| \leq 1/2$ and $0 \leq \rho_0 \leq 1$. Writing
$\rho_j(\xi,\eta) = \rho(\xi/j,\eta/j)$ ($j=1,2,\ldots$) we obtain
$\supp(\rho_j) \subseteq [-j,j]^2$ and $\rho_j \to 1$ uniformly over
compact subsets of $\R^2$ as $j\to \infty$. Hence
\[
   w_{x,t} = \D'-\lim\limits_{j\to\infty} \int\limits_{-j}^j 
   \int\limits_{-j}^j e^{i\phi_t(.;\xi,\eta)} a_j(\xi,\eta)\, d\xi
   d\eta  = \D'-\lim\limits_{j\to\infty} w^j_{x,t}
\]
where 
\[
   a_j(\xi,\eta) =  \rho_j(\xi,\eta) a(\xi,\eta) = \rho_j(\xi,\eta)
   \frac{\chi(\xi) \chi(\eta/\ga)}{\ga |\xi| |\eta|} .
\]

Since $\supp(a_j) \subseteq [-j,j]^2$ is compact $s \mapsto
w^j_{x,t}(s)$ is smooth and by differentiating inside the integral we
see that for all $l\in\N_0$ $(\diff{s})^l w^j_{x,t}(s)$ is bounded by
some constant (depending on $l$ and $a_j$). Hence
$(w^j_{x,t})_{j\in\N}$ is a sequence in the space $\S'(\R)$ of
tempered distributions. Therefore, to prove that $w_{x,t}$ is in
$\S'(\R)$, it suffices to show that $(w^j_{x,t})_{j\in\N}$ converges
weakly in $\S'(\R)$, i.e., for all rapidly decaying smooth functions
$\vphi\in\S(\R)$ the sequence $\dis{w^j_{x,t}}{\vphi}$ is convergent.

We have 
\begin{eqnarray*}
   \dis{w^j_{x,t}}{\vphi}
   &=& \int \vphi(s)\!\!\int e^{\msci \phi_t(.;\xi,\eta)} 
            a_j(\xi,\eta)\, d(\xi,\eta)\, ds
\\
   &=& \int e^{-\msci (t|\eta| + x(\xi -\eta/\ga))} a_j(\xi,\eta) 
   \!\!\int\!\! e^{\msci s(|\xi| - |\eta|)}
                                      \vphi(s)\, ds\; d(\xi,\eta)
\\
   &=& \int e^{-\msci (t|\eta| + x(\xi -\eta/\ga))} a_j(\xi,\eta) 
       \FT{\vphi}(|\eta| - |\xi|) \, d(\xi,\eta) .  
\end{eqnarray*}
Here, the integrand tends pointwise to
 $e^{-\msci (t|\eta| + x(\xi -\eta/\ga))} a(\xi,\eta)
  \FT{\vphi}(|\eta| - |\xi|)$ as $j\to\infty$
and is dominated by $|a(\xi,\eta) \FT{\vphi}(|\eta| - |\xi|)|$. It
remains to show that $(\xi,\eta) \mapsto a(\xi,\eta) \FT{\vphi}(|\eta|
- |\xi|)$ is in $\L^1(\R^2)$; then an application of Lebesgue's
dominated convergence theorem will provide us with existence of an
explicit integral expression for the limit $\dis{w_{x,t}}{\vphi}$.

Since $\FT{\vphi}\in\S(\R^2)$, using the explicit structure of $a$, we
have for any $k\in\N$ a bound of the form
\[
   |a(\xi,\eta) \FT{\vphi}(|\eta| - |\xi|)| \leq 
   C_k (1+|\xi|)^{-1} (1+|\eta|)^{-1} (1+||\eta| - |\xi||)^{-k} 
   \ \ \forall (\xi,\eta)\in\R^2 .
\]
While integrating the right-hand side of this inequality over $\R^2$,
we split the integration into four parts according to the sign
combinations of $\xi$ and $\eta$. By symmetry, this boils down to
estimating only the two kinds of integrals
\[
   I_- = \int\limits_0^\infty \int\limits_0^\infty 
         \frac{d\xi d\eta}{(1+\xi)(1+\eta)(1+|\eta-\xi|)^k}, \quad 
   I_+ = \int\limits_0^\infty \int\limits_0^\infty
         \frac{d\xi d\eta}{(1+\xi)(1+\eta)(1+\eta+\xi)^k} . 
\]
In $I_+$ we only have to note that $(1+\xi+\eta)^{-k} \leq
(1+\xi)^{-k/2} (1+\eta)^{-k/2}$ which together with the remaining
factors gives a finite integral as soon as $k > 0$. In $I_-$ we change
variables to $\nu = \eta - \xi$, $\mu = \eta$ to obtain
\[
   I_- = \int\limits_{-\infty}^\infty \frac{1}{(1+|\nu|)^k} \!\!
         \int\limits_{\max(0,\nu)}^\infty \!\! 
         \frac{d\mu}{(1+\mu)(1+\mu-\nu)} \, d\nu .
\]
In the inner integral we use $1 + \mu - \nu = (1+\mu)(1 - \nu/(1+\mu))
\geq (1+\mu)/(1+|\nu|)$ yielding an upper bound $(1+|\nu|)
\int_0^\infty (1+\mu)^{-2} d\mu$ and hence
\[
   I_- \leq \int\limits_0^\infty \frac{d\mu}{(1+\mu)^2} 
            \int\limits_{-\infty}^\infty \frac{d\nu}{(1+|\nu|)^{k-1}}
\]
which is finite if $k > 2$. This proves the assertion that $(\xi,\eta)
\mapsto a(\xi,\eta) \FT{\vphi}(|\eta| - |\xi|)$ is indeed in
$\L^1(\R^2)$ and establishes the following result.
\begin{prop} $w_{x,t}\in\S'(\R)$ and for any $\vphi\in\S(\R)$ 
\begin{equation}\label{w_action}
   \dis{w_{x,t}}{\vphi}
   = \lim\limits_{j\to\infty} \dis{w^j_{x,t}}{\vphi}
   = \int e^{-\msci(t|\eta| + x(\xi -\eta/\ga))} a(\xi,\eta) 
          \FT{\vphi}(|\eta| - |\xi|) \, d(\xi,\eta) .
\end{equation}
\end{prop}

We are now in a position to determine the Fourier transform of
$w_{x,t}$ explicitly.
\begin{prop} $\FT{w_{x,t}}$ is the continuous function on $\R$ given
by (the classical integral)
\begin{equation}\label{FT_w}
   \FT{w_{x,t}}(r)
   = 4\pi\, e^{itr}\!\! \!\!\int\limits_{\{|\xi| \geq r\}}\!\!\!\!  
     e^{-\msci (x\xi + t|\xi|)} \cos(\frac{x}{\ga}(|\xi|-r))
     a(\xi,|\xi|-r)\, d\xi .
\end{equation}
\end{prop}
\begin{proof}
Let $\vphi\in\S(\R)$ then $\FT{\FT{\vphi}}(s)= 2\pi \vphi(-s)$ and
from (\ref{w_action}) we obtain
\begin{eqnarray*}
   \dis{\FT{w_{x,t}}}{\vphi} &=& \dis{w_{x,t}}{\FT{\vphi}}
\\
   &=& 2 \pi \int e^{-i(t|\eta| + x(\xi -\eta/\ga))} a(\xi,\eta) 
         \vphi(|\xi| - |\eta|) \, d(\xi,\eta)
\\
   &=& 2 \pi \int e^{-ix\xi} \Big( 
         \int\limits_{-\infty}^0 
         e^{-\msci (-t\eta - x\eta/\ga)} a(\xi,\eta)
           \vphi(|\xi| + \eta)\, d\eta
\\
   & &   \hphantom{2 \pi \int e^{-ix\xi}} 
         + \int\limits_0^\infty 
         e^{-\msci (t\eta - x\eta/\ga)} a(\xi,\eta)
           \vphi(|\xi| - \eta)\,d\eta \Big)\, d\xi ,
\end{eqnarray*}
where in the last line we have made use of the symmetry properties of
$a(\xi,\eta)$. Changing coordinates in the inner integrals to $r =
|\xi| \pm \eta$ and again by the symmetry of $a(\xi,\eta)$ this reads
\[
   2\pi \int e^{-\msci (x\xi + t|\xi|)} \int\limits_{-\infty}^{|\xi|}
   e^{\msci t r} a(\xi,|\xi|-r) \vphi(r) 
   \underbrace{(e^{i(r-|\xi|)x/\ga} - e^{-i(r-|\xi|)x/\ga})}_{2
               \cos(\frac{x}{\ga}(|\xi|-r))} \, dr \, d\xi .
\]
Finally, since $\vphi\in\S$ and $|a(\xi,|\xi|-r)| \leq p(r)
(1+|\xi|)^{-2}$ for some polynomial in $r$, we may interchange the
order of integration and arrive at
\[
   \dis{\FT{w_{x,t}}}{\vphi}
   = \int \vphi(r) \cdot 4\pi e^{itr} \!\!\!\! 
     \int\limits_{\{|\xi| \geq r \}} \!\!\!\! e^{-i(x\xi + t|\xi|)} 
     \cos(\frac{x}{\ga}(|\xi|-r)) a(\xi,|\xi|-r)\, d\xi \, dr .
\]
Since $\vphi$ was arbitrary and the above upper bound for
$a(\xi,|\xi|-r)$ shows that the inner integrand is in $\L^1$ w.r.t.
$\xi$, the proposition is proved.
\end{proof}

From (\ref{FT_w}) we immediately obtain the correlation by setting
$c(t) = \FT{w_{x,t}}(0)$, in the form
\begin{eqnarray}
   c(t) &=& 4\pi \int\limits_{-\infty}^\infty 
            e^{-\msci (x\xi + t|\xi|)}
            \cos\left(\frac{x\xi}{\ga}\right)
            a(\xi,\xi)\, d\xi  
\label{corr_int}\\
        &=& \frac{2\pi}{\ga} \Big( \int\limits_{-\infty}^\infty
            e^{-\msci (x\xi(1-1/\ga) + t|\xi|)} 
            \frac{\chi(\xi)\chi(\frac{\xi}{\ga})}{|\xi|^2}\, d\xi
\nonumber\\
        & & \phantom{\frac{2\pi}{\ga} \Big(}
            - \int\limits_{-\infty}^\infty 
            e^{-\msci (x\xi(1+1/\ga) + t|\xi|)}  
            \frac{\chi(\xi)\chi(\frac{\xi}{\ga})}{|\xi|^2}\, d\xi
            \Big) .
\nonumber
\end{eqnarray}
This shows that $t \mapsto c(t)$ is continuous and can be represented
as the difference of two (classically convergent) OIs with symbols of
order $-2$, and hence $c \in \L^1(\R)$. Note that the (distributional)
derivative $c'(t)$ can be obtained by differentiating w.r.t.\ $t$
inside the OI raising the order of the symbol by one. Therefore $c'$
will not be continuous on the whole line.

Finally, we observe that again the information about the singularity
shift is revealed by the singular support of $c(t)$. By the
stationarity condition on the phase functions, we find
\[
   \WF(c) \subseteq \{\pm x(1+1/\ga), \pm x(1-1/\ga)\} \times \R_+ ,
\]
where $\pm x(1-1/\ga)$ represent the true shifts from $\pm x/\ga$ to
$\pm x$ whereas $\pm x(1+1/\ga)$ are the distances from $\mp x/\ga$ to
$\pm x$. It is easily seen that $c(t)$ cannot be smooth at the points
$t = \pm x(1 \pm 1/\ga)$, e.g., by noting that each time derivative
brings down a new factor of $|\xi|$ in each integrand, and at the $t$
values in question one of the phase functions vanishes identically
along a half-line in $\xi$. Hence, we have in fact the exact
information
\begin{equation}\label{singsupp_c_3}
   \singsupp(c) = \{\pm x(1+1/\ga), \pm x(1-1/\ga)\} ,
\end{equation}
which also fits nicely with remark \ref{weak_smooth_remark} on the
weak smoothness of $w_{x,t}$.

\section{Microlocalization of the correlation}

From the case studies, we conjecture that the singular support of the
correlation of two wave fields reveals the relative shift in wave
front sets between them. As we pointed out, in general, the critical
point set of the correlation need not be compatible with this
shift. Here, we propose an alternative approach to extract the shift
from the correlation, viz., by detecting its singular support. We
design a pseudodifferential operator that enables this detection. Our
approach can be applied invariably to any derivative of the
correlation also.

In the generic case, the correlation $c \in \S'$ with Fourier
transform $\FT{c}$. Let $\phi$ be the Gaussian in one dimension,
define
\beq
   \psi_{r,t}(s) = \frac{1}{r} \phi\left( \fr{s - t}{r} \right) .
\eeq
Introduce
\beq
   W_{\psi_{r,t}} c(\tau) = \FT{\psi_{r,t}} \ast \FT{c}(\tau)
   \quad\mbox{for}\quad \tau = \pm 1 ,
\eeq
a continuous wavelet transform that can be written as the action of a
pseudodifferential operator $\psi_{r,t}(D_{\tau})$ (in ${\mathrm{Op}}\ 
  S^{-\infty}$) on $\FT{c}$. The growth properties reveal the wave
  front set at $t$ in the direction $\pm 1$. In fact, $(t,\pm 1)
  \not\in \WF{c}$ if for any $N \in \N$,
\beq
   |W_{\psi_{r,t}} c(\pm 1)| \le C_N r^N
   \quad\mbox{for}\quad r \in ]0,1]
\eeq
(see \cite{FO:89}). Effectively, this leads to a scanning procedure
over $t$: whenever the condition is not satisfied, $t \in
\singsupp(c)$. In particular, this applies if
 $|W_{\psi_{r,t}} c(\pm 1)| \approx r^M$
for some fixed $M$.

If $c$ would allow an OI representation, as is the case in the
examples of Section~4, we could apply a stationary phase argument
instead, as in (\ref{wf_osc}).

\section{Discussion}

Starting from the microlocal representation, we analyzed the
measurement process of wave fields. Such process can be described by a
restriction operator. We then adressed the issue of how the detection
of wave front sets propagates through the measurement process. Then we
focused on the detection of (base) shifts in wave front sets due to
perturbation of the wave field within the measurement. We introduced
the distributional cross-correlation as a tool for this purpose, and
analyzed its properties. In a series of case studies, we investigated
in what way the cross-correlation reveals the shifts. In the first
case the correlation was a measure, in the second case it was a
bounded measurable function, and in the third case it was a continuous
function. It was conjectured that the time shift coincides with the
singular support of the correlation. We proposed a procedure (a
pseudodifferential operator) to detect the shift based on
microlocalization. Such procedure would comprise the foundation for
wave-equation tomography.

\paragraph{Acknowledgement:} We thank J\'{e}r\^{o}me Le Rousseau for
valuable mathematical remarks and improvements of the text.



\newcommand{\SortNoop}[1]{}

\end{document}